\documentclass[conference]{IEEEtran}
\ifCLASSINFOpdf
\else
\fi
\usepackage{amsmath}
  \usepackage[caption=false,font=normalsize,labelfont=sf,textfont=sf]{subfig}

\usepackage{url}
\usepackage{hyperref}
\usepackage{listings}
\usepackage{color}
\usepackage{tikz}
\usetikzlibrary{arrows,automata,positioning,trees,shapes,
decorations.pathreplacing,trees}

\usepackage{tkz-graph}

\begin{document}
%
\title{Counterexample-Guided \textit{k}-Induction for Fast Bug Detection}

\author{\IEEEauthorblockN{Mikhail Y. R. Gadelha}
\IEEEauthorblockA{Electronics and Computer Science,\\
University of Southampton, \\
Southampton, UK \\
Email: myrg1g14@soton.ac.uk}
\and
\IEEEauthorblockN{Lucas C. Cordeiro}
\IEEEauthorblockA{Department of Computer Science,\\
University of Oxford,\\
Oxford, UK\\
Email: lucas.cordeiro@cs.ox.ac.uk}
\and
\IEEEauthorblockN{Denis A. Nicole}
\IEEEauthorblockA{Electronics and Computer Science,\\
University of Southampton, \\
Southampton, UK \\
Email: dan@ecs.soton.ac.uk}
}


%


\maketitle

\begin{abstract}
Recently, the \textit{k}-induction algorithm has proven to be a successful
approach for both finding bugs and proving correctness. However, since the
algorithm is an incremental approach, it might waste resources trying to prove
incorrect programs. In this paper, we propose to extend the \textit{k}-induction
algorithm in order to shorten the number of steps required to find a property
violation. We convert the algorithm into a meet-in-the-middle bidirectional
search algorithm, using the counterexample produced from over-approximating the
program. The preliminary results show that the number of steps required to find
a property violation is reduced to $\lfloor\frac{k}{2} + 1\rfloor$ and the
verification time for programs with large state space is reduced considerably.
\end{abstract}


%
\IEEEpeerreviewmaketitle

\section{Introduction}

Embedded systems are used in a variety of applications, ranging from nuclear
plants and automotive systems to entertainment and games~\cite{heath}. This
ubiquity drives a need to test and validate a system before releasing it to the
market, in order to protect against system failures. Even subtle system
bugs can have drastic consequences, such as the recent Heartbleed bug on
OpenSSH, which might have leaked private information from several
servers~\cite{Durumeric:2014:MH:2663716.2663755}.

One promising technique to validate a system is called bounded model checking
(BMC)~\cite{Biere:1999:SMC:646483.691738}. The basic idea of BMC is to check
the negation of a property at a given depth: given a transition system $M$, a
property $\phi$, and a bound $k$, BMC unrolls the system $k$ times and
generates verification conditions (VC) $\psi$, such that $\psi$ is satisfiable
if and only if $\phi$ has a counterexample of depth $k$ or less.  BMC tools
based on Boolean Satisfiability (SAT) or Satisfiability Module Theories (SMT)
have been applied on the verification of both sequential and parallel
programs~\cite{Clarke04, CordeiroFM12, qadeer05}. However, BMC tools are
aimed to find bugs; they cannot prove correctness, unless the bound $k$ safely
reaches all program states.

Despite the fact that BMC cannot prove correctness by itself (unless it fully
unwinds the program), there are algorithms that use BMC as a ``component'' to
prove correctness. In particular, the \textit{k}-induction algorithm is an
incremental algorithm that aims to find bugs and prove correctness using an
ever increasing number of unwindings. In this paper, we propose to extend the
algorithm originally developed for \textit{k}-induction to shorten the number
of iterations required to find a property violation. Our main original contribution
is an extension to the \textit{k}-induction algorithm, which
converts the algorithm into a meet-in-the-middle bidirectional search by
using the counterexample generated by the inductive check
(Section~\ref{extension}). The preliminary results show that the number of
steps required to find a property violation is reduced to
$\lfloor\frac{k}{2} + 1\rfloor$ and the verification time for programs with large
state space is reduced considerably (Section~\ref{pre-res}).

\section{The \textit{k}-induction Algorithm}
\label{kinduction}

The first version of the \textit{k}-induction algorithm was proposed by
Sheeran et al.~\cite{Sheera00}; they apply BMC to find bugs and prove
correctness. BMC tools cannot prove correctness unless the bound $k$ is
appropriate to reach the completeness threshold (i.e., a value that will fully
unroll all loops occurring in the program, often impractically
large~\cite{Kroening2011}). For instance, consider the simple program shown in
Figure~\ref{figure:unwindingprograms1}, the assertion in
line~\ref{code-example:assert} always holds, regardless of the initial value of
\texttt{n} in line~\ref{code-example:init}. BMC tools as CBMC~\cite{Clarke04},
ESBMC~\cite{CordeiroFM12} or LLBMC~\cite{MerzFS12}
typically produce the program in Figure~\ref{figure:unwindingprograms2} and are
unable to verify that program unless the loop is fully unrolled, i.e., the
\textit{unwinding assertion} if $k < 2^{32}-1$ in $32$-bit and $64$-bit
architectures.
\begin{figure}[h]
\centering
\begin{minipage}{.9\columnwidth}
    \begin{lstlisting}[escapechar=^]
int main() {
  uint32_t n;^\label{code-example:init}^
  uint64_t sn = 0;
  for (uint64_t i = 1; i <= n; i++) {
    sn = sn + 2;
    assert(sn == i * 2);
  }
  assert(sn == n*2 || sn == 0);^\label{code-example:assert}^
}
    \end{lstlisting}
\end{minipage}
  \caption{Simple loop program.}
  \label{figure:unwindingprograms1}
\end{figure}

In mathematics, one usually approaches such class of problems using
\textit{proof by induction}. The \textit{k}-induction variant has been
successfully combined with continuously-refined invariants~\cite{Beyer15}, was
used to prove that C programs do not contain data races~\cite{Donaldson10},
or that design time constraints are respected~\cite{EenS03}; it is a
well-established technique in hardware verification, where it is applied due to
the monolithic transition relation present in hardware
designs~\cite{Sheera00,GrosseLD09}.

\begin{figure}[h]
\centering
\begin{minipage}{.9\columnwidth}
  \begin{lstlisting}[escapechar=^]
int main() {
  uint32_t n;
  uint64_t sn = 0;
  uint64_t i = 1;
  if(i <= n) {
    sn = sn + 2;        ^\raisebox{-9pt}[0pt][0pt]{\setlength\delimitershortfall{-17pt}$\Bigg\}\:k\text{~copies}$}^
    assert(sn == i * 2);
    i++;
  }
  assert(!(i<=n)); // unwinding assertion
  assert(sn == n*2 || sn == 0);
}
    \end{lstlisting}
  \end{minipage}
  \caption{Finite $k$ unwindings done by BMC.}
  \label{figure:unwindingprograms2}
\end{figure}

We define the \textit{k}-induction algorithm as an iterative deepening search
algorithm~\cite{Russell:2003:AIM:773294}. Let a given program under
verification $P$ be a finite transition system $m$ with branching factor
$b$ and depth $d$ (we use $b^+$ to represent the overapproximated branch factor,
such that $b \subseteq b^+$, and $d^+$ to represent the overapproximated depth,
such that $d \subseteq d^+$), $I(s_n)$ and $T(s_n, s_{n+1})$ be the formulae
for the initial states and transition relations for $m$ over propositional state
variables $s_n$ and $s_{n+1}$,
$\Phi$ be the set of safety properties, $\phi(s) \in \Phi$ be the formula
encoding for states satisfying a safety property and $\psi(s)$ be the formula
encoding for states satisfying the completeness threshold. We also define a
counterexample as a sequence of states $[s_1,\ldots,s_k]$ of length $k$ that
violates a safety property. The \textit{k}-induction algorithm is a complete
and optimal (always find the shortest counterexample) search algorithm, with
complexity $O(bd)$ and state space $O(b^+d^+)$. Jovanovi\'{c} et
al.~\cite{Jovanovic:2016:PK:3077629.3077648} prove that \textit{k}-induction
can be more powerful and concise than regular induction.

In each step $k$ of the \textit{k}-induction algorithm, three checks are
performed: the base case $B(k)$, forward condition $F(k)$ and inductive step
$I(k)$, for $k=[1,d]$~\cite{Gadelha2015}. The base case $B(k)$ is the
standard BMC and $B(k)$ is satisfiable if and only if $B(k)$ has a
counterexample of length \textit{k} or less~\cite{handbook09}:
\begin{equation}\label{eq:bk}
  B(k) = I(s_1) \wedge \bigvee^k_{i=1} \bigwedge^{i-1}_{j=1} T(s_j, s_{j+1})
  \wedge \neg \phi(s_i).
\end{equation}

The second check, the forward condition $F(k)$, checks if the
completeness threshold $\psi(s)$ holds for the current $k$. This is established
by checking if the following is unsatisfiable:
\begin{equation}\label{eq:fk}
  F(k) = I(s_1) \wedge \bigvee^k_{i=1} \bigwedge^{i-1}_{j=1} T(s_j, s_{j+1})
  \wedge \neg \psi(s_k).
\end{equation}

No safety property $\phi(s)$ is checked in $F(k)$ as they were
already checked for the current \textit{k} in the base case. Finally, the
inductive step $I(k)$, checks if whenever $\phi(s)$ holds in \textit{k}
states $s_1,\ldots,s_k$, $\phi(s)$ also holds for the next state $s_{k+1}$.
This is established by checking if the following is unsatisfiable:
\begin{equation}\label{eq:ik}
  I(k) = \bigwedge^{k-1}_{j=1} T(s_j, s_{j+1}) \wedge \phi(s_i)
  \wedge \neg \phi(s_k).
\end{equation}

Combining the three checks, the \textit{k}-induction algorithm at a
given $k$ is:
\begin{equation}\label{eq:verk}
  kind(P, k) =
  \begin{cases}
    P \text{ contains a bug}, & \text{if}\ \neg B(k) \\
    P \text{ is correct}, & \text{if}\ F(k) \vee I(k) \\
    kind(P, k+1), & \text{otherwise}.
  \end{cases}
\end{equation}

\section{Extending the \textit{k}-induction Algorithm}
\label{extension}

The \textit{k}-induction algorithm is being applied to solve a number of
different verification problems, but the algorithm has limitations, that our work
addresses. The biggest limitation is the fact that if a state $\xi$
violates a property at depth $k$, the algorithm requires $k$ steps to find
the counterexample. This is expensive because of the three checks performed for
each $k$. The inductive check is the most computationally expensive
of the three checks; it is an overapproximation, forcing the SMT solver to
find a set of assignments in a larger state space than the original
program~\cite{Gadelha2015}. Moreover, the computation is wasted if a
counterexample is found by the inductive check, as it is assumed to be spurious.

In order to address the problem of the wasted computation and of the \textit{k}
iterations, we propose to use the counterexample generated by
the inductive check, to speed up the bug finding check (the base case). By
reusing the counterexample found by the inductive check, we aim to cut the
number of required \textit{k} steps to find a bug in half.

The main idea is to search simultaneously both forward (from the initial state
$s_1$) and backward (from the error state $\xi$) and stop if the two
searches meet in the middle. Assuming that the error state is reachable in $k$
steps from the initial state, the solution will be found in
$\lfloor\frac{k}{2} + 1\rfloor$, because the forward and backward searches each
have to go only half way ($b$ is the branching
factor)~\cite{Russell:2003:AIM:773294}.

Our extension aims to convert the \textit{k}-induction algorithm into a
bidirectional search algorithm, by using the base case as the forward part and
the inductive step as the backward part. The algorithm will still be complete
and optimal, and the state space explored is much smaller than $O(b^+d^+)$ as
the number of states and transition relations evaluated by each step is smaller.
Note that the extension repurposes the goal of the inductive check, from
proving correctness to find paths that lead to error states.
\begin{figure}[h]
  \begin{minipage}{.5\textwidth}
    \centering
    \begin{tikzpicture}
\SetVertexMath
\GraphInit[vstyle=Empty]

\Vertex[L=s_1]{A}
\Vertex[x=-1,y=-1,L=s_2]{B}
\Vertex[x=1,y=-1,L=s_3]{C}
\Vertex[x=-2,y=-2,L=s_4]{D}
\Vertex[x=0,y=-2,L=s_5]{E}
\Vertex[x=2,y=-2,L=s_6]{F}

\Vertex[x=0,y=-3,L=\vdots]{S}

\Vertex[x=-2,y=-4,L=s_{n-9}]{G}
\Vertex[x=0,y=-4,L=s_{n-8}]{H}
\Vertex[x=1,y=-4,L=s_{n-7}]{I}
\Vertex[x=-1,y=-5,L=s_{n-6}]{L}
\Vertex[x=2,y=-5,L=s_{n-5}]{M}
\Vertex[x=-2,y=-6,L=s_{n-4}]{J}
\Vertex[x=0,y=-6,L=s_{n-3}]{O}
\Vertex[x=1,y=-6,L=s_{n-2}]{P}
\Vertex[x=2,y=-6,L=\xi]{Q}
\Vertex[x=-1,y=-7,L=s_{n-1}]{N}
\Vertex[x=1,y=-7,L=s_n]{R}

\tikzset{EdgeStyle/.style={->}}
\Edges(A,B,D)
\Edges(A,B,E)
\Edges(A,C,F)
\Edges(G,J)
\Edges(L,J)
\Edges(H,L,O,N)
\Edges(H,P)
\Edges(I,M,Q)
\Edges(P,R)

\Vertex[x=3.5,y=0]{k=1}
\Vertex[x=3.5,y=-1]{k=2}
\Vertex[x=3.5,y=-2]{k=3}

\Vertex[x=3.5,y=-4]{k=d-3}
\Vertex[x=3.5,y=-5]{k=d-2}
\Vertex[x=3.5,y=-6]{k=d-1}
\Vertex[x=3.5,y=-7]{k=d}

\draw [dashed,decorate] ([yshift=-.5cm, xshift=5.5cm]A.west)--([yshift=-.5cm,
xshift=-3cm]A.east) node {};

\draw [dashed,decorate] ([yshift=-1.5cm, xshift=5.5cm]A.west)--([yshift=-1.5cm,
xshift=-3cm]A.east) node {};

\draw [dashed,decorate] ([yshift=-2.5cm, xshift=5.5cm]A.west)--([yshift=-2.5cm,
xshift=-3cm]A.east) node {};

\draw [dashed,decorate] ([yshift=-3.5cm, xshift=5.5cm]A.west)--([yshift=-3.5cm,
xshift=-3cm]A.east) node {};

\draw [dashed,decorate] ([yshift=-4.5cm, xshift=5.5cm]A.west)--([yshift=-4.5cm,
xshift=-3cm]A.east) node {};

\draw [dashed,decorate] ([yshift=-5.5cm, xshift=5.5cm]A.west)--([yshift=-5.5cm,
xshift=-3cm]A.east) node {};

\draw [dashed,decorate] ([yshift=-6.5cm, xshift=5.5cm]A.west)--([yshift=-6.5cm,
xshift=-3cm]A.east) node {};

\draw [|->,decorate] ([xshift=4.5cm]A.east)--([xshift=2.5cm]F.east) node
[yshift=1cm, right] {\textit{B(k)}};

\draw [|->,decorate] ([xshift=2.5cm]Q.east)--([xshift=3.3cm]I.east) node
[yshift=-1cm, right] {\textit{I(k)}};
\end{tikzpicture}
  \end{minipage}
  \caption{Visual representation of our proposed extension. Each dashed section
represents the states reachable after $k$ iterations. The arrows show the
``direction'' of the verification by the base case $B(k)$ and the inductive
check $I(k)$. The forward condition $F(k)$ is not shown in this representation
but it is a forward check, similar to $B(k)$.}
  \label{figure:extension}
\end{figure}
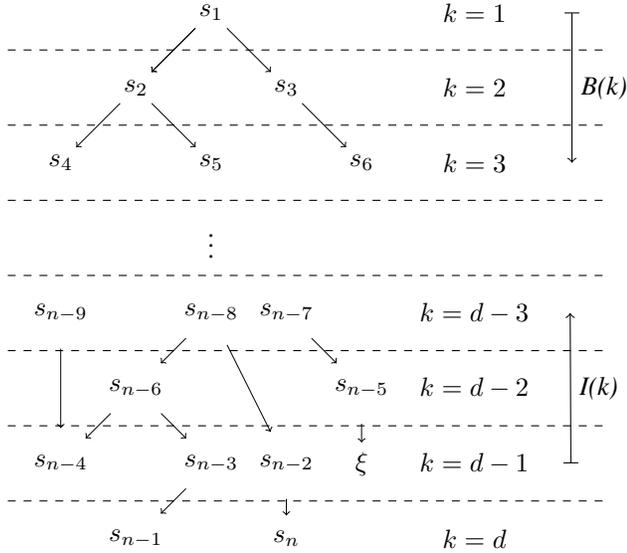

Figure~\ref{figure:extension} is a visual representation of the proposed
extension, for a given set of states $s_1\ldots s_n$ and an error state $\xi$.
In this representation, each \textit{k} step of the \textit{k}-induction
algorithm verifies up to $k$ levels in the graph.

The base case is the forward part of the algorithm, it tries to find a
counterexample [$s_1$,\ldots,$\xi$] of length $k$. The inductive check is the
backward part of the algorithm; as defined by Equation~\ref{eq:ik}, the
inductive check tries to find a counterexample $[s,\ldots,\xi]$ of length $k$,
from anywhere in the graph.

The counterexample produced by the inductive check is then a path that leads to
a property violation; if at least one state of this path is reachable from the
initial state $s_1$, then the error state $\xi$ is reachable from the initial
state $s_1$. Given a counterexample $[s,\ldots,\xi]$ from the inductive check,
our extension translates that into a new safety property $\varphi(s_i, s)$:
\begin{equation}\label{eq:prop-ind}
 \varphi(s_i, s) =
  \begin{cases}
    1, & \text{if}\ s_i = s \\
    0, & \text{otherwise}.
  \end{cases}
\end{equation}

Note that we do not check if a given state $s_i$ is in a counterexample
$[s,\ldots,\xi]$, but rather check if the given state $s_i$ is the first state
of the counterexample. Given the optimal nature of the algorithm, this is
sufficient to find the property violation. The new safety property is then
checked in the new base case $B'(k)$:
\begin{equation}\label{eq:bik}
  B'(k) = I(s_1) \wedge \bigvee^k_{i=1} \bigwedge^{i-1}_{j=1} T(s_j, s_{j+1})
  \wedge \neg \phi(s_i) \wedge \neg \varphi(s_i).
\end{equation}

The computational complexity added to the \textit{k}-induction algorithm
by our extension is minimal, as it only adds new property checks to the base
case. The inductive check, the most computationally expensive check, remains
unaltered by our extension. The extended \textit{k}-induction is also sound: if
the counterexample generated by the inductive is spurious, no state in the
sequence will be reachable from the initial state.

\section{Experimental Evaluation}
\label{pre-res}

In order to evaluate the extension to the \textit{k}-induction algorithm, we
selected a number of benchmarks from the International Competition on Software
Verification (SV-COMP) 2017~\cite{svcomp2017}. We compare the results from the
original \textit{k}-induction and our extended version. The programs evaluated
by our extension were manually changed to add the invariants generated by the
inductive step.

\subsection{Description of the Benchmarks}
\label{Description-of-the-Benchmarks}

The benchmarks called sum0* are similar to the program in
Figure~\ref{figure:unwindingprograms1}, but contain a bug in different depths.
The benchmarks rangesum* check if a function is ``deterministic'' w.r.t.
all possible permutations of an input array; the number in the benchmark name
represents the size of the array. The benchmark const\_false checks if a constant
holds after 1024 iterations (but checks the wrong value after the iterations);
diamond checks if a counter that is being nondeterministically incremented is even
after 99 iterations; and Problem01\_label15 is the representation of a reactive
system.

All experiments were conducted on a computer with an Intel Core i7-2600
running at 3.40GHz and 24GB of RAM under Fedora 25 64-bit. We used ESBMC
v4.5~\cite{CordeiroFM12} and no time or memory limit were set for the verification tasks.

\subsection{Objectives}
\label{objectives}

\begin{enumerate}

\item[RQ1] \textbf{(performance)} does the extended \textit{k}-induction
require less resources than the original \textit{k}-induction, w.r.t. time and
memory?

\item[RQ2] \textbf{(sanity)} Does the extended \textit{k}-induction provide any
incorrect result?

\end{enumerate}

\subsection{Results}
\label{results}

Table~\ref{table:pre-res} shows the preliminary results obtained from the
original \textit{k}-induction and our proposed extension. Here, $L$ is the number
of lines in the program, $Time$ is the time needed to verify the
program in seconds, $Mem$ is the memory used by the tools to verify the programs
in megabytes\footnote{We used the command \texttt{/usr/bin/time -v} from linux to
measure both the time and the memory usage} and $k$ is the number of steps
needed to find the bug. The last lines show the average and cumulative numbers
for each of the columns. We order the benchmarks in relation to the memory
required by the original \textit{k}-induction. The best results are marked in
bold.

\begin{table*}[t!]
\begin{center} {\small
\begin{tabular}{|l|r||r|r|c|r|r|c|}
\hline
Benchmark &  & \multicolumn{3}{c|}{\parbox{3.1cm}{\centering\textit{k}-induction}}   & \multicolumn{3}{c|}{Extended \textit{k}-induction} \\
\cline{3-8}
                                         & $L$   &$Time$ (s)&$Mem$ (MB)& $k$  &$Time$(s)    &$Mem$(MB) & $k$   \\
\hline
sum04\_false-unreach-call.c              & 19    &\textbf{1}&\textbf{38.7}& 9 &\textbf{1}   &\textbf{38.7}&\textbf{6}   \\
\hline
sum01\_false-unreach-call.c              & 18    &\textbf{1}& 38.9     & 11   &\textbf{1}   &\textbf{38.8}&\textbf{6}   \\
\hline
sum03\_false-unreach-call.c              & 25    & 3        & 39.1     & 11   &\textbf{1}   &\textbf{38.8}&\textbf{6}   \\
\hline
diamond\_false-unreach-call1.c           & 24    & 13       & 43.6     & 51   &\textbf{6}   &\textbf{39.1}&\textbf{26}  \\
\hline
rangesum\_false-unreach-call.c           & 64    & 7        & 66.2     & 4    &\textbf{1}   &\textbf{39.0}&\textbf{2}   \\
\hline
rangesum05\_false-unreach-call.c         & 59    & 11       & 72.3     & 6    &\textbf{1}   &\textbf{65.4}&\textbf{3}   \\
\hline
rangesum10\_false-unreach-call.c         & 59    & 28       & 78.2     & 11   &\textbf{16}  &\textbf{47.5}&\textbf{6}   \\
\hline
Problem01\_label15\_false-unreach-call.c & 594   & 7        & 87.3     & 5    &\textbf{5}   &\textbf{70.3}&\textbf{4}   \\
\hline
rangesum20\_false-unreach-call.c         & 59    & 101      & 99.9     & 21   &\textbf{26}  &\textbf{78.2}&\textbf{12}  \\
\hline
rangesum40\_false-unreach-call.c         & 59    & 847      & 269.5    & 41   &\textbf{90}  &\textbf{113.9}&\textbf{22}  \\
\hline
const\_false-unreach-call1.c             & 20    & 2606     & 796.6    & 1024 &\textbf{890} &\textbf{253.2}&\textbf{513} \\
\hline
rangesum60\_false-unreach-call.c         & 59    & 80272    & 1106.9   & 61   &\textbf{159} &\textbf{134.6}& \textbf{32}  \\
\hline\hline
Average                                  & 88    & 6991     & 228.1    & 104  & 99          & 79.79        & 53 \\
\hline
Total                                    & 1059  & 83897    & 2737.2   & 1255 & 1197        & 957.5        & 638 \\
\hline
\end{tabular} }
\end{center}
\caption{Preliminary results over the SV-COMP benchmarks.}
\label{table:pre-res}
\end{table*}

The first noticeable aspect of the results is that the time of the verification
is not related to the number of steps or the program size. The closest relation
between the verification time is the state space explored by each step
(more specifically the inductive check), the bigger the state space, longer it
will take to find a solution; this can somehow be summarized by memory used by
the tool during the verification.

The evaluation for this set of benchmarks show that our extension to the
\textit{k}-induction algorithm potentially cuts the verification time
considerably in cases where the state space explored is large. For small cases
(e.g., the sum0*\_false-unreach-call.c benchmarks), our extension does not slow
down or uses more memory than the original \textit{k}-induction and for large
cases, the gains were substantial (the verification time of
rangesum60\_false-unreach-call.c was 504x faster). In terms of
the steps needed to find the bug, the extended version of the
\textit{k}-induction required $\lfloor\frac{k}{2} + 1\rfloor$, as expected.

For each benchmark, the verification time and memory usage is either equal or
smaller, compared to the original \textit{k}-induction, and thus affirm RQ1.
Regarding the results (known to contain bugs as they are part of the SV-COMP),
the extended \textit{k}-induction provided the same results that the original
\textit{k}-induction provided, positively answering RQ2.

\section{Related Work}
\label{related}

The extension proposed in this paper is basically a form of target enlargement,
where a target state $s$ is ``enlarged'' by precomputing the set of states that
may hit $s$ in $k$-steps. Here, we analyse two works in that direction.

Bischoff et al.~\cite{BISCHOFF200533} propose a methodology to use BDDs and
SAT solvers for the verification of programs. The BDDs are responsible for the
target enlargement, collecting the under-approximate reachable state sets,
followed by the SAT-based verification with the newly computed sets. The authors
implemented the technique in the Intel BOolean VErifier (BOVE) and showed that
the time was up to five times smaller. Compared to this work, we only use
\textit{k}-induction and SMT solvers; the inductive check in the
\textit{k}-induction algorithm is responsible for enlarging the target and the SMT
solver checks for satisfiability.

Jovanovi\'{c} et al.~\cite{Jovanovic:2016:PK:3077629.3077648} present a
reformulation of IC3, separating the reachability checking from the inductive
reasoning. They further replace the regular induction algorithm by the
\textit{k}-induction algorithm and show that it provides more concise
invariants. The authors implemented the algorithm in the SALLY model checker
using Yices2 to do the forward search and MathSAT5 to do the backward search.
They showed that the new algorithm is able to solve a number of real-world
benchmarks, at least as fast as other approaches. Compared to this work, our
proposed extended \textit{k}-induction uses consequent BMC calls to find a
solution. We also implement our approach independent of solvers and it can be
used with any SMT solver supported by ESBMC; however, both searches will be
done with the same solver.

\section{Conclusion}
\label{conclusions}

In this paper, our main contribution is a novel extension to the
\textit{k}-induction algorithm, to perform a bidirectional search instead of
the conventional iterative deepening search. The extension is currently under
development using ESBMC. We plan to evaluate the improvement over the SV-COMP
benchmarks, where the original \textit{k}-induction algorithm already proved to
be the state-of-art, if compared to other \textit{k}-induction
tools~\cite{svcomp2017}.

The preliminary results show that the extension has the potential to
substantially improve the verification time for problems with large state space,
while maintaining a small verification time for small programs. In one
particularly large program (in terms of state space), our extension allowed
the \textit{k}-induction algorithm to find the property violation on average
using half of the steps and a fraction of the resources.





%

\end{document}